\documentclass[twocolumn,showpacs,prb,aps,epsf,amsmath,amssymb]{revtex4}

\usepackage{graphicx}

\begin{document}

\title{Donor binding energy and thermally activated persistent photoconductivity in high mobility (001) AlAs quantum wells}

\author{S.~Dasgupta, C.~Knaak, J.~Moser, M.~Bichler, S.F.~Roth, A.~Fontcuberta i Morral, G.~Abstreiter}
\affiliation{
\centerline{Walter Schottky Institut, Technische Universit\"at M\"unchen, Garching, D-85748 Germany}  
}
\author{M.~Grayson\footnote{Corresponding author: m-grayson@northwestern.edu}} 
\affiliation{
\centerline{Walter Schottky Institut, Technische Universit\"at M\"unchen, Garching, D-85748 Germany}
\centerline{and}
\centerline{Department of Electrical Engineering and Computer Science, Northwestern University, Evanston, IL 60208 USA}
}

\begin{abstract}

A doping series of AlAs (001) quantum wells with Si $\delta$-modulation doping on both sides reveals different dark and post-illumination saturation densities, as well as temperature dependent photoconductivity. The lower dark two-dimensional electron density saturation is explained assuming deep binding energy of $\Delta_{\mathrm {DK}} =$ 65.2~meV for Si-donors in the dark. Persistent photoconductivity (PPC) is observed upon illumination, with higher saturation density indicating shallow post-illumination donor binding energy. The photoconductivity is thermally activated, with 4~K illumination requiring post-illumination annealing to $T$ = 30~K to saturate the PPC. Dark and post-illumination doping efficiencies are reported. 

\end{abstract}


\maketitle

Two dimensional electron systems (2DESs) in aluminum arsenide (AlAs) quantum wells (QWs) are interesting for their valley degeneracy and heavy mass \cite{gunawan,shkolnikov,shayegan}.  The valley index quantum number acts as an extra pseudospin degree of freedom, and the heavy mass allows interactions to play a larger role at a given density \cite {depoor3}.  Recently progress has also been made in fabricating and characterizing one-dimensional AlAs nanostructures \cite{moser1,moser2,vakili}. Although improvements in high mobility AlAs 2DES structures have been reported \cite{depoor2,depoor4}, many important material parameters such as the donor binding energy and doping efficiency have been obscured by substrate charge effects \cite{depoortere}. Since these parameters are instrumental in designing and optimizing heterostructures, we have performed a systematic study on double-sided-doped quantum wells which screen away unwelcome substrate effects.  In the process, we also identify a thermally activated persistent photoconductivity (PPC) not previously reported. 

AlAs is an indirect band gap III-V semiconductor with three degenerate conduction band valleys at the X-points of the Brillouin zone edge. In (001) growth, the biaxial strain between AlAs and Al$_x$Ga$_{1-x}$As ($x=0.45$) decreases the energy of the two in-plane valleys such that for wide wells $W >$ 55 \AA,\cite{vanKest,vandest} only these two valleys are degenerately occupied. In AlAs, the longitudinal and transverse electron masses are anisotropic, $m_{\mathrm l}$ = 1.1 $m_{\mathrm e}$ and $m_{\mathrm t}$ = 0.2 $m_{\mathrm e}$, respectively \cite {shayegan}, and the effective Land\'e $g$-factor $g^*$ = 2 (Ref. 11).

Free electrons in AlAs/AlGaAs heterostructures come from two different sources: intentional Si-dopant and unintentional dilute charge traps. Firstly, when AlGaAs is intentionally doped with Si, the Si can act as a substitutional donor supplying a hydrogenically bound electron, or it can act as a DX-center which captures electron charge until illuminated \cite{schubert,mooney}. At an aluminum content $x$ = 0.45, the hydrogenic binding energies from a charged subsitutional impurity can be either shallow, when associated with the light-mass $\Gamma$ band, or deep, when associated with the heavier degenerate X,$\Gamma$ and L bands \cite{bourgoin,Shklovskii}. The second source of space-charge is the unintentional dilute charge traps in the high $x$-content Al$_x$Ga$_{1-x}$As barriers \cite{depoortere,schubert,fischer,roth}.  A field effect persistent photoconductivity in single-side-doped AlAs wells has been reported \cite{depoortere}, whereby the electron density in the 2DES depends on the strength of a gate bias during illumination.  It was proposed that in this case the electrons in the QW come only partly from the top donor layer of Si, and the remainder come from dilute charge traps in the AlGaAs barrier below the quantum well \cite{depoortere}. But the uncertainties about the intrinsic doping inherent in the charge trap density leave it impossible to separate out doping efficiencies and donor binding energies of the Si dopants in such structures. Furthermore, the front side doped AlAs QWs do not conduct in the dark \cite{depoortere}. In our experiments, we have screened out the effects of dilute substrate charge with an additional backside doping and systematically deduced the donor binding energy and doping efficiency from the saturated dark and post-illumination density of a doping series of AlAs QWs. Unlike previous work, our samples conduct in the dark and do not require gating to work reproducibly after illumination, and our best sample has slightly better mobility than the best reported sample at comparable densities and temperatures \cite{depoor2}.  We identify an unreported PPC mechanism for the Si-donors which saturates only after heating to 30 K, and define an illumination protocol to reproducibly achieve maximum density and mobility which we call post-illumination annealing (PIA).  

\begin{figure}
	\includegraphics[width=\linewidth,keepaspectratio]{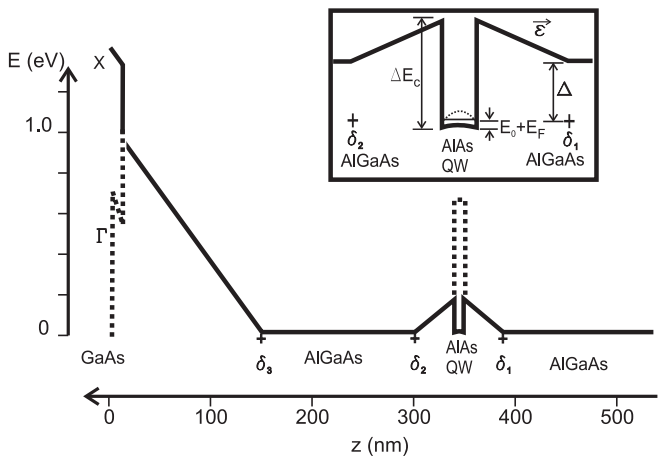}
	\caption{(a) Energy band diagram of the AlAs QW. The z-axis denotes the growth direction in nm and the vertical axis shows the energy scale in eV. The + denotes the Si delta doping layers in the AlGaAs layers. The solid line shows the X-point band and the dashed line shows the $\Gamma$-point band. (b) Inset depicts the QW in detail with electric field across the spacer $\cal E$, donor binding energy $\Delta$, conduction band offset $\Delta E_{\mathrm C}$, confinement energy $E_0$ and Fermi energy $E_{\mathrm F}$ relative to the well minimum.}
	\label{fig:graph1}
\end{figure}

The samples were grown on (001) GaAs substrates using molecular beam epitaxy. The structure of the sample shown in Fig.~1, is a 150 {\AA} wide QW with three Si $\delta$ doping layers, grown with a Si current of 11.4 A. $\delta_1$ and $\delta_2$ separated from the QW by Al${_{0.45}}$Ga${_{0.55}}$As spacers provide electrons to the QW. The spacer $d$ is 350 {\AA} on the surface side and 450 {\AA} on the substrate side. The dopant layers are placed asymmetrically to take into account an estimated 50 {\AA} forward diffusion of Si during growth\cite{lanz} resulting in symmetric $d$ = 400 {\AA} spacers as modeled in Fig.~1. These two delta doping layers are doped equally with a Si density $n_{\delta_1}$ = $n_{\delta_2}$ = $n_{\mathrm {Si}}$. Since the nominally undoped AlGaAs layers are known to have dilute charge traps \cite{depoortere,schubert,fischer,roth} the backside doping layer $\delta_1$ is crucial to pin the substrate potential relative to $E_{\mathrm F}$ upon saturation, effectively screening out these substrate charges. The Si doping near the surface with three times the density $n_{\delta_3}$ = $3n_{\mathrm {Si}}$, satisfies charge traps at the sample surface and pins the conduction band above the quantum well to $E_{\mathrm F}$ upon saturation. Various samples were grown with different doping $n_{\mathrm {Si}}$ indexed A-F \cite{sample} in Fig.~3(b).

\begin{figure}
	\includegraphics[width=\linewidth,keepaspectratio]{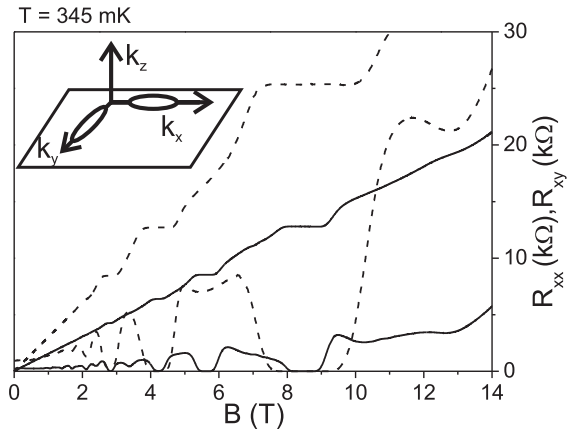}
  \caption{$R_{\mathrm {xx}}$ and $R_{\mathrm {xy}}$ measurements at 345 mK dark (dashed) and post-illumination (solid) for (001) sample C. Due to the high transport mass of AlAs, the quantum Hall features are seen only below 1 K. In the dark, $n = 2.0\times 10^{11}$ cm$^{-2}$ and $\mu$ = $1.3\times 10^{5}$ cm$^{2}$/Vs and post-illumination, $n = 4.0\times 10^{11}$ cm$^{-2}$ and $\mu$ = $2.7\times 10^{5}$ cm$^{2}$/Vs. Inset shows the two occupied valleys.}
	\label{fig:graph2}
\end{figure}

Indium contacts made by hand with indium solder, with a contact area of roughly $1.0$ mm$^{2}$, were annealed at 450 $^{\circ}$C for 100 s. Two-point contact resistance was around 100 k$\Omega$ at 300 K and around 40 k$\Omega$ at 4.2 K. Samples were illuminated using an infra-red (IR, 950 nm) or red (635 nm) LED. Transport measurements were performed on van der Pauw samples and L-shaped Hall bars down to 330 mK. Mobilities were found to be isotropic to within 1.2 \%. Typical longitudinal ($R_{\mathrm {xx}}$) and transverse ($R_{\mathrm {xy}}$) resistance with magnetic field at 345 mK in the dark and post-illumination for sample C is plotted in Fig.~2. The density of the samples in subsequent figures were deduced from such measurements. 

We observe an unanticipated thermal activation of the PPC. When the samples are illuminated at 4 K with an IR or red LED, the density of the samples may remain unaffected or increase by a factor of up to 1.5 depending upon intensity. But when the low temperature illumination is turned off and the sample is heated up to 30~K, the density then doubles from the dark value. This experiment has been plotted in Fig.~3(a). The density starts from its 4 K post-illumination value of about $2.0\times 10^{11}$ cm$^{-2}$ and increases around 25 K, to a maximum of about $4.0\times 10^{11}$ cm$^{-2}$ and remains unchanged upon recooling to 4.2 K.

We therefore define a new illumination protocol to include a subsequent anneal to 30 K, and refer to this as "`post-illumination anneal"' (PIA). This standard illumination protocol was used for all samples to reproducibly achieve the maximum density and mobility for the data reported in Figs.~2 and 3(b).

\begin{figure}
 \includegraphics[width=\linewidth,keepaspectratio]{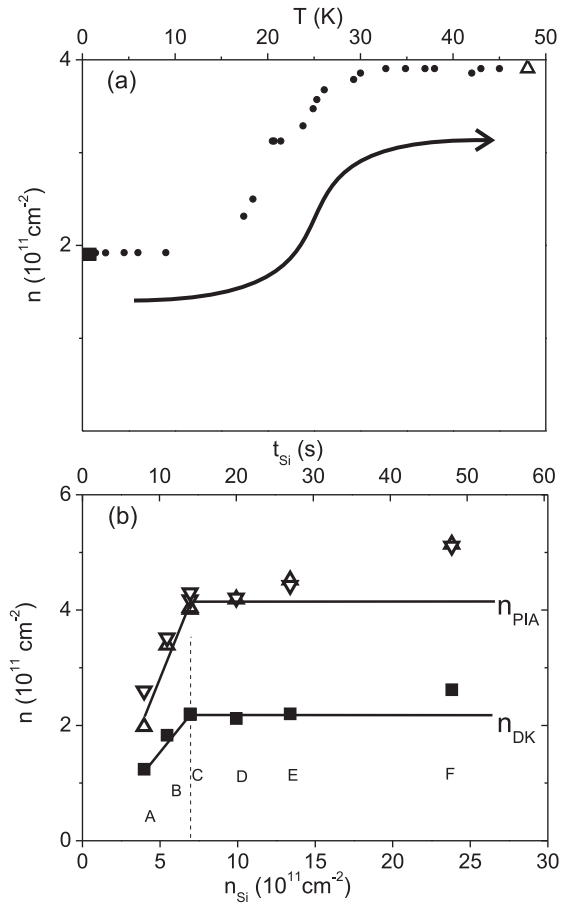}
 \caption{(a) Illustration of post illumination thermal anneal(PIA) for sample C. Post illumination when the sample is returned to dark at low T, density remains the same as $n_{\mathrm {DK}}$($\blacksquare$). When T is increased up to 30 K, the saturation density $n_{\mathrm {PIA}}$ is obtained($\bigtriangleup$). The arrow shows the time progression. (b) The density of two dimensional electron system in an (001) oriented AlAs quantum well as a function of the Si delta doping $n_{\mathrm {Si}}$. The top $x$-axis defines the time of Si doping corresponding to the doping densities. The vertical dashed line shows the saturation threshold $n_{\mathrm{sat}}$ for both the dark and the post illumination(red LED $\triangledown$, IR LED $\bigtriangleup$) conditions at 1.4 K. The solid lines are plotted as an aid to locating the saturation threshold.}
	\label{fig:graph4}
\end{figure}

In Fig.~3(b) the 2DES density of the samples are plotted as a function of different Si doping concentrations $n_{\mathrm {Si}}$ deduced from the doping times $t_{\mathrm {Si}}$ and a calibrated Si flux of $4.9\times 10^{10}$ cm$^{-2}$ s$^{-1}$ for 11.4 A Si heater current. The dark electron density is seen to saturate at $n_{\mathrm {DK}}$ = $2.0\times 10^{11}$ cm$^{-2}$. Post-illumination anneal (PIA) density, plotted as triangles, is seen to saturate at about $n_{\mathrm {PIA}}$ = $4.0\times 10^{11}$ cm$^{-2}$. Sample C represents the optimally doped sample, since at higher doping approximiately the same saturation density $n_{\mathrm {PIA}}$ is observed, and it also has the highest mobility of all samples, $\mu_{DK} = 1.3\times 10^5$ cm$^2$/Vs and $\mu_{PIA} = 2.7\times 10^5$ cm$^2$/Vs. We note that whereas in GaAs samples excess modulation dopant leads to a parallel conduction channel in the dark, in this higher Al-content AlGaAs the increased effective mass of electrons in the deep traps correspondingly decreases the Bohr radius, preventing hopping conduction from occurring in this parallel dopant layer. 

We can deduce the binding energies of the donors by studying the electrostatics represented in Fig.~1. Solving the one-dimensional Poisson equation, we obtain the electric field $\cal{E}$ in the spacer to be 
\begin{equation}
\label{equ:electricfield}
	{\cal E} = \frac{\Delta E_c - (E_0 + E_{\mathrm F}) - \Delta}{ed} = \frac{en_{\mathrm {sat}}}{2 \epsilon_{0} \epsilon_{\mathrm r}} .
\end{equation}
\newline The conduction band mismatch between AlAs and Al${_{0.45}}$Ga${_{0.55}}$As is $\Delta E_{\mathrm C}$ = 140 meV \cite{vurf}. From Hartree simulations, we calculate the binding energy of the ground state in the well to be $E_0$ = 12 meV (dark) and 14 meV (PIA) relative to the X-band minimum at the AlAs/AlGaAs interface (using $m_\mathrm{t}$ as the mass in the confinement direction). The Fermi energy is $E_{\mathrm {F}}$ = 0.53 meV for the dark saturation density of $2.0\times 10^{11}$ cm$^{-2}$ and 1.06 meV for the PIA saturation density of $4.0\times 10^{11}$cm$^{-2}$. $\epsilon_0$ denotes the permittivity of free space and the relative permittivity of Al$_{0.45}$Ga$_{0.55}$As is $\epsilon_{\mathrm r}$ = 11.6 \cite{dielectric}. The density-of-states mass used for Fermi energy is $m^* =(m_{\mathrm l} m_{\mathrm t})^{1/2} = 0.45 m_{\mathrm e}$. Because the spacer thickness $d$ and the other parameters listed above are all known, the donor binding energy $\Delta$ can be deduced from Eq.~1 after measuring the dark or post-illumination saturation density $n_{\mathrm {sat}}$.

Solving Eq.~(1) for $\Delta$, the experimental dark density of $n_{\mathrm {DK}}$ yields a donor binding energy of $\Delta_{\mathrm {DK}} = $65.2~meV.  This agrees with binding energies of bulk Si-doped Al$_{0.45}$Ga$_{0.55}$As which are shown to be of order 100 meV near the $\Gamma$-X-L-band degeneracy \cite{chand}. Applying the same analysis to $n_{\mathrm {PIA}}$ to deduce the post-illumination saturation binding energy, we arrive at $\Delta_{\mathrm {PIA}} = 0.0$ meV as an estimate of the shallow donor bound state relative to the X-conduction band.  Given the uncertainty in the dielectric constant of Al$_{0.45}$Ga$_{0.55}$As, as well as the dilute charge traps in the spacer, this value may reasonably represent a shallow bound state of a few meV.  We speculate that this shallow bound state may be associated with hydrogenically bound $\Gamma$-band electrons.  One could also assume that the $\Gamma$-band sits a few meV above the X-band at $x = 0.45$, so the observed $\Delta_\mathrm{PIA} = 0.0$ meV relative to the X-band may represent a reasonable binding energy a few meV below the $\Gamma$-band.

The doping efficiency for the samples in the dark and post-illumination was calculated using the doping density $n_{\mathrm {Si}}$ at the saturation threshold. We define
\newline ~\begin{equation}
\label{equ:doping efficiency}
	\eta_{\mathrm {DK, PIA}} = \frac{n_{\mathrm {DK, PIA}}}{2 n_{\mathrm {Si}}}
\end{equation}
\newline where $\eta$ is the doping efficiency. The factor of 2 in the denominator arises from the twin top and bottom $\delta$ doping layers. Using this equation, we obtain the doping efficiency of $\eta_{\mathrm {DK}}=16\%$ in the dark and $\eta_{\mathrm {PIA}} = 31\%$ post-illumination anneal.  The increased doping efficiency after illumination proves that some electrons are optically activated from a dark bound state, and indicative of DX centers. This structure can only calibrate the doping efficiency exactly at the saturation density since at higher doping densities the 2D density saturates, and at lower $n_{\mathrm {Si}}$, the band structure can be affected by unscreened electric fields in the substrate of unknown magnitude.

We can now propose a phenomenological model to explain the dark and PIA saturation densities.  Upon cooling the sample from room temperature, the quantum well is populated with electrons from the deepest hydrogenic donor state, and some electrons remain bound on the donor atoms which act as DX centers.  Upon illumination, the DX-bound electrons are excited to higher energy metastable states\cite{mooney,bourgoin,chand, saxena}, which they can leave only via thermal excitation.  At temperatures from 15-25 K electrons are then excited from the metastable state and fall either into the QW or into shallow hydrogenic impurity levels in the donor layer. If these metastable levels are more likely to relax to this shallow $\Gamma$-band-related impurity state in the donor layer than to the deeply bound impurity state, a higher post-illumination saturation density would result, according to Eq.~1.  When thermally cycled above 50~K, there is an indication that the density starts to decrease again, presumably due to electrons rebinding to DX-centers.

In summary, we have been able to deduce the binding energy of Si in $\delta$-doped layers in Al${_{0.45}}$Ga${_{0.55}}$As in the dark and after illumination. This sample design is shown to work in the dark, and still yield a mobility slightly better than previously reported samples which do not show dark conduction. The dark Si dopants appear to populate both deeply bound states which pin the Fermi energy well below the conduction band, as well as DX-centers which capture the some of the electron charge.  Upon illumination, the DX-centers then transfer their electrons to metastable bound states. When the sample temperature is raised up to 30 K these metastable states release their charge to the QW or to shallow hydrogenically bound donor states resulting in twice the free carrier saturation density in the QW. The doping efficiency for the Si $\delta$-layers has been calculated. These parameters will be instrumental in optimizing mobility in future AlAs QWs. 
\\
\\
This work was funded by BmBF Nano-Quit Project 01BM470. The authors would also like to thank Dieter Schuh, Jens M\"uller and Stefan Birner for discussions.


\begin{thebibliography}{10}
\bibitem{dielectric}      
   $\epsilon_r$ is a linear interpolation of the low temperature $\epsilon_{\mathrm {GaAs}}$ = 12.7 \cite{samara} and $\epsilon_{\mathrm {AlAs}}$ = 10.16 \cite{guyaux} according to: $\epsilon_{\mathrm {AlGaAs}} = (1-x) \epsilon_{\mathrm {GaAs}} + (x) \epsilon_{\mathrm {AlAs}}$.
\end{thebibliography}
\end{document}